# Experimental realization of a topological crystalline insulator in SnTe

Y. Tanaka[1], Zhi Ren[2], T. Sato[1], K. Nakayama[1], S. Souma[3], T. Takahashi[1,3], Kouji Segawa[2], & Yoichi Ando[2].

[1]*Department of Physics, Tohoku University, Sendai 980-8578, Japan*

[2]*Institute of Scientific and Industrial Research, Osaka University, Ibaraki, Osaka 567-0047, Japan*

[3]*WPI Research Center, Advanced Institute for Materials Research, Tohoku University, Sendai 980-8577, Japan*

**Topological insulators materialize a topological quantum state of matter where unusual gapless metallic state protected by time-reversal symmetry appears at the edge or surface[1,2]. Their discovery stimulated the search for new topological states protected by other symmetries[3-7], and a recent theory[8] predicted the existence of "topological crystalline insulators" (TCIs) in which the metallic surface states are protected by mirror symmetry of the crystal. However, its experimental verification has not yet been reported. Here we show the first and definitive experimental evidence for the TCI phase in tin telluride (SnTe) which was recently predicted to be a TCI[9]. Our angle-resolved photoemission spectroscopy shows clear signature of a metallic Dirac-cone surface band with its Dirac point slightly away from the edge of the surface Brillouin zone in SnTe. On the other hand, such a gapless surface state is absent in a cousin material lead telluride (PbTe), in line with the theoretical prediction. Our result establishes the presence of a TCI phase, and opens new avenues for exotic topological phenomena.**

The surface state of three-dimensional (3D) topological insulators (TIs) is characterized by a spin non-degenerate Dirac-cone energy dispersion protected by time-reversal symmetry (TRS). In TIs, the TRS plays a key role in characterizing the topological properties such as the quantum spin Hall effect, the dissipation-less spin current, and the magnetoelectric effect[10-13]. In contrast, in TCIs metallic surface states are protected by *mirror symmetry* of the crystal[8]. Recent tight-binding calculations[9] predicted a narrow-gap IV-VI semiconductor SnTe to be a TCI owing essentially to its intrinsically inverted band structure and mirror symmetry of the face-centered-cubic (f.c.c.) lattice of rocksalt structure (see *e.g.*, Fig. 1a), while the isostructural PbTe is not a TCI because of the absence of the band inversion. It is thus of particular importance to experimentally examine the possibility of the TCI phase in these semiconductors, in order to establish the concept of this new topological state of matter and possibly to find novel topological phenomena beyond the framework of known topological materials.

In our angle-resolved photoemission spectroscopy (ARPES) experiment, we paid particular attention to the momentum space around the $\bar{\text{X}}$ point of the surface Brillouin zone (BZ) corresponding to a projection of the L point of the bulk BZ (Fig. 1a) where a direct bulk-band gap resides[14-16] and the appearance of topological surface states are predicted[9]. A mirror plane that characterizes the TCI properties lies in the (110) plane, which is projected to the $\bar{\Gamma}\bar{\text{X}}$ high-symmetry line in the surface BZ. As shown in Fig. 1b, the ARPES intensity at the Fermi level ($E_F$) measured with the photon energy $h\nu =$ 21.2 eV on the (001) surface exhibits bright intensity pattern centered around the $\bar{\text{X}}$ point and is elongated along the $\bar{\Gamma}\bar{\text{X}}$ direction. The band dispersion along two selected cuts (red arrows in Fig. 1b) exhibits a linearly dispersive holelike feature crossing $E_F$, as displayed in Figs. 1c and d. The top of this Dirac-like band is located not at the $\bar{\text{X}}$ point but at a point slightly away from it (called here the $\bar{\Lambda}$ point), as one can see in the band dispersion along the $\bar{\Gamma}\bar{X}$ cut (Figs. 1e and f) showing the band maxima in both sides of the $\bar{X}$ point ($\bar{\Lambda}_1$ and $\bar{\Lambda}_2$ for the first and second surface BZs, respectively). Such a characteristic "M"-shaped dispersion is not expected from the bulk-band calculation at any $k_z$ (momentum perpendicular to the surface) values[14-16], suggesting that the observed Dirac-like band originates from the surface states.

To further examine whether the Dirac-like band is of surface or bulk origin, we have carried out an ARPES measurement along the cut crossing the $\bar{\Lambda}$ point for various photon energies. As one immediately recognizes from Figs. 1g-j, energy position of the band is stationary with respect to the $h\nu$ variation. In fact, when we plot the extracted dispersions for different photon energies in the same panel, they overlap each other within the experimental uncertainties for the binding energy ($E_B$) of less than ~0.2 eV (Fig. 1m), demonstrating the surface origin of the Dirac-like band. We emphasize here that, since as-grown crystals of SnTe tend to show heavily hole-doped nature[17,18], a key to the present observation of the Dirac-like band was to reduce extra hole carriers in the crystal by minimizing Sn vacancies during the sample fabrication. In fact, the Dirac-like surface state was not resolved in the previous ARPES study[16] mainly due to heavily hole-doped nature of the sample (chemical potential was located ~ 0.5 eV lower compared to our data).

As shown in Fig. 1k, the ARPES data at $T$ = 130 K divided by the Fermi-Dirac (FD) distribution function present that the left- and right-hand-side dispersion branches actually merge into a single peak at above $E_F$. The Dirac-point energy is estimated to be 0.05 eV above $E_F$ from a linear extrapolation of the two dispersion branches (Fig.

1m). Also, one can see in Fig. 1m that the band dispersion exhibits no discernible change with temperature (compare 30-K and 130-K data for $h\nu$ = 92 eV); the Dirac band velocity extracted from the dispersions are 4.5 and 3.0 eVÅ, for left- and right-hand-side branches, respectively. This signifies that the small rhombohedral distortions along the (111) direction that occurs below 100 K (ref. 19), which would break the mirror symmetry with respect to the (110) plane, are not strong enough to disturb the Dirac-like surface band dispersion at low temperature.

To further elucidate the topology of the Dirac-like band in detail, we have determined the whole band dispersion in the two-dimensional (2D) momentum space. By selecting a specific photon energy ($h\nu$ = 92 eV), thanks to the matrix-element effects, we found it possible to pick up the dispersion of a single Dirac cone centered at the $\bar{\Lambda}_2$ point ($\bar{\Lambda}_2$ Dirac cone) while suppressing the intensity of the $\bar{\Lambda}_1$ Dirac cone. Figures 2a-e show near-$E_F$ ARPES intensity measured along several cuts (A-E) around the $\bar{\Lambda}_2$ point. Along cut A (Fig. 2a), the holelike surface band has its top at $E_B$ of 0.45 eV. Upon moving from cut A to E, the band maximum (white arrow) approaches $E_F$ (cuts A-B), passes $E_F$ (cut C), and then disperses back again toward higher $E_B$ (cuts D-E). This result establishes the cone-shaped dispersion of the Dirac-like band in the 2D momentum space like in 3D TIs[1,2] and graphene[20]. In passing, we have surveyed electronic states throughout the BZ and found no evidence for other metallic surface states (see, *e.g.* Fig. 1b) and thus concluded that the surface electronic states consist of four Dirac cones in the first surface BZ. This indicates that this material is not a TI due to an even number of band-inversion points[9].

Unlike prototypical 3D TIs[1,2], the Dirac cone in SnTe is anisotropic and slightly elongated along the $\overline{\Gamma X}$ direction, as is visible in the ARPES-intensity contour plots for several $E_B$ slices shown in Fig. 2g. This is reasonable because the $\overline{\Lambda}$ point is not a high-symmetry point in the BZ and has only mirror symmetry with respect to the $\overline{\Gamma X}$ line, as is also highlighted in the 3D band dispersion derived from the present experiment (Fig. 2h). Interestingly, such an elongated shape of the Dirac cone at the off-symmetry point is also observed in the antiferromagnetic phase of an iron-pnictide $BaFe_2As_2$ (ref. 21), although the origin of the Dirac cone is distinctly different between the two cases.

One may ask if the appearance of the Dirac-cone surface states is a unique feature of SnTe among isostructural IV-VI semiconductors. To address this point, we have performed an ARPES measurement of PbTe and directly compared the near-$E_F$ electronic states around the $\overline{\Lambda}$ point, as displayed in Figs. 3a and b. Intriguingly, the spectral feature of PbTe shows no evidence for the metallic Dirac-like band, and displays a broad feature originated from the bulk valence band[22], suggesting that this material is an ordinary (trivial) insulator. This naturally suggests that a topological phase transition takes place in a solid-solution system $Pb_{1-x}Sn_xTe$ (Fig. 3c). This conclusion is also supported by a tight-binding calculation[9] which predicted that the valence bands at four L points in SnTe are inverted relative to PbTe, resulting in different mirror Chern numbers (2 *vs*. 0). One can thus infer that the bulk-band gap closes at a critical $x$ value, $x_c$, accompanied by a parity change of the valence-band wave function and an emergence/disappearance of the Dirac-cone surface state. Therefore, the present results have established for the first time the TCI phase in SnTe, which is in contrast to a trivial nature of isostructural PbTe. Our results unambiguously demonstrate the validity of the concept of TCI and suggests the existence of many more kinds of topological materials.

**METHODS**

High-quality single crystals of SnTe and PbTe were grown in sealed evacuated quartz-glass tubes from high-purity elements [Sn (99.99%), Pb (99.998%), Te (99.999%)]. To obtain SnTe crystals with minimal Sn vacancy, a starting ratio of Sn:Te=51:49 was chosen and, after melting the elements at high temperature, the tube was slowly cooled to 770°C (which is only 20°C below the melting point) and quenched into cold water; the carrier density estimated from the room-temperature Hall coefficient is $2 \times 10^{20}$ cm$^{-3}$, which is an order of magnitude smaller than that in ordinary SnTe crystals[16]. ARPES measurements were performed with the MBS-A1 and VG-Scienta SES2002 electron analyzers with a high-intensity helium discharge lamp at Tohoku University and also with tunable synchrotron lights at the beamline BL28A at Photon Factory (KEK). To excite photoelectrons, we used the He Iα resonance line ($h\nu$ = 21.218 eV) and the circularly polarized lights of 50-100 eV at Tohoku University and Photon Factory, respectively. The energy and angular resolutions were set at 10-30 meV and 0.2°, respectively. Samples were cleaved *in-situ* along the (001) crystal plane in an ultrahigh vacuum of $1 \times 10^{-10}$ Torr. A shiny mirror-like surface was obtained after cleaving the samples, confirming their high quality. The Fermi level of the samples was referenced to that of a gold film evaporated onto the sample holder.

**REFERENCES**

1. Hasan, M. Z. & Kane, C. L. Colloquium: Topological insulators. *Rev. Mod. Phys.* **82**, 3045-3067 (2010).

2. Qi, X.-L. & Zhang, S.-C. Topological insulators and superconductors. *Rev. Mod. Phys.* **83**, 1057-1110 (2011).

3. Schnyder, A. P., Ryu, S., Furusaki, A. & Ludwig, A. W. W. Classification of topological insulators in three spatial dimensions. *Phys. Rev. B* **78**, 195125 (2008).

4. Kitaev, A. Periodic table for topological insulators and superconductors. Preprint at <http://arxiv.org/abs/0901.2686v2> (2009).

5. Ran, Y. Weak indices and dislocations in general topological band structures. Preprint at <http://arxiv.org/abs/1006.5454v2> (2010).

6. Mong, R. S. K., Essin, A. M. & Moore, J. E. Antiferromagnetic topological insulators. *Phys. Rev. B* **81**, 245209 (2010).

7. Li, R., Wang, J., Qi, X.-L. & Zhang, S.-C. Dynamical axion field in topological magnetic insulators. *Nature Phys.* **6**, 284-288 (2010).

8. Fu, L. Topological crystalline insulators. *Phys. Rev. Lett.* **106**, 106802 (2011).

9. Hsieh, T. H., Lin, H., Liu, J., Duan, W., Bansil, A. & Fu, L. Topological crystalline insulators in the SnTe material class. Preprint at <http://arxiv.org/abs/1202.1003> (2012).

10. Kane, C. L. & Mele, E. J. $Z_2$ topological order and the quantum spin Hall effect. *Phys. Rev. Lett.* **95**, 146802 (2005).

11. Bernevig, B. A., Hughes, T. L. & Zhang, S.-C. Quantum spin Hall effect and topological phase transition in HgTe quantum wells. *Science* **314**, 1757-1761 (2006).
7

**Acknowledgements**

We thank Liang Fu for stimulating discussions. We also thank M. Komatsu, M. Nomura, E. Ieki, N. Inami, H. Kumigashira, and K. Ono for their assistance in ARPES measurements, and T. Ueyama and K. Eto for their assistance in crystal growth. This work was supported by JSPS (NEXT Program and KAKENHI 23224010), JST-CREST, MEXT of Japan (Innovative Area "Topological Quantum Phenomena"), AFOSR (AOARD 124038), and KEK-PF (Proposal number: 2012S2-001).

**Author Contributions**

Y.T., T.S., K.N., S.S., and T.T. performed ARPES measurements. Z.R., K.S., and Y.A. carried out the growth of the single crystals and their characterizations. Y.T., T.S., and Y.A. conceived the experiments and wrote the manuscript.

**Competing Interests Statement**

We have no competing financial interests.

**Correspondence** and requests for materials should be addressed to T.S. (e-mail: t-sato@arpes.phys.tohoku.ac.jp) or Y.A. (e-mail: y_ando@sanken.osaka-u.ac.jp).

**FIGURE LEGENDS**

**FIG. 1 | Dirac-like band dispersion in SnTe.** **a**, Bulk BZ (red lines) and corresponding (001) surface BZ (blue lines). (110) mirror plane is indicated by green shaded area. **b**, ARPES intensity mapping at $E_F$ at $T$ = 30 K for SnTe plotted as a function of 2D wave vector measured at $h\nu$ = 21.2 eV; this intensity is obtained by integrating the spectra within ±10 meV of $E_F$. **c** and **d**, Near-$E_F$ ARPES intensity measured at $h\nu$ = 21.2 eV as a function of wave vector and $E_B$ along the cut crossing the $\bar{\Lambda}_1$ and $\bar{\Lambda}_2$ point (red arrows in **b**), respectively. **e** and **f**, Energy distribution curves (EDCs) along the $\bar{\Gamma}\bar{X}$ cut (yellow arrow in **b**) measured at $h\nu$ = 21.2 eV, and corresponding intensity plot, respectively. Dashed lines in **e** are a guide to the eyes to trace the band dispersion. **g**-**j**, ARPES intensity measured at $T$ = 30 K with various photon energies across the cut crossing the $\bar{\Lambda}_2$ point (green arrow in **b**). **k**, The same as **g** but measured at $T$ = 130 K. The ARPES intensity is divided by the FD distribution function convoluted with the instrumental resolution. **l**, Slice of bulk BZ in the (110) plane, together with the momentum points in which the ARPES data for **c**-**d** and **g**-**k** were obtained. $k_z$ values were estimated by using the inner-potential value of 8.5 eV as determined by the normal-emission ARPES measurement. **m**, Comparison of the band dispersion for various photon energies extracted by tracking the peak position of momentum distribution curves.

**FIG. 2 | Two-dimensional band dispersion of SnTe.** **a-e**, Near-$E_F$ ARPES intensity for SnTe as a function of wave vector and $E_B$ measured at $h\nu$ = 92 eV along the cuts (A-E) around the $\bar{\Lambda}_2$ point shown by red arrows in the surface BZ in **f**. White arrow marks the top of dispersion. **g**, ARPES intensity mappings for SnTe as a

function of 2D wave vector at various $E_B$'s. Note that the intensity distribution below 0.45 eV shows a topology distinctly different from that of the Dirac cone since it originates from the bulk valence band. **h**, 3D plots of near-$E_F$ band dispersion obtained by tracing the peak position of EDCs around the $\bar{\Lambda}$ point. The energy band is folded with respect to the $\bar{X}\bar{X}$ line by taking into account the symmetry of the BZ. Blue circles show energy contours for representative $E_B$ slices, highlighting the Lifshitz transition[9].

**FIG. 3 | Comparison of band structure between SnTe and PbTe.** **a**, Comparison of near-$E_F$ EDCs around the $\bar{\Lambda}$ point between SnTe and PbTe measured with $h\nu$ = 21.2 eV. Note that the energy axes are different for SnTe and PbTe. **b**, Corresponding ARPES-intensity plots. **c**, Schematic illustration of the evolution of the band dispersion in Pb$_{1-x}$Sn$_x$Te expected from the present ARPES experiment. SS, CB, and VB denote the surface state, the bulk conduction band, and the bulk valence band, respectively.

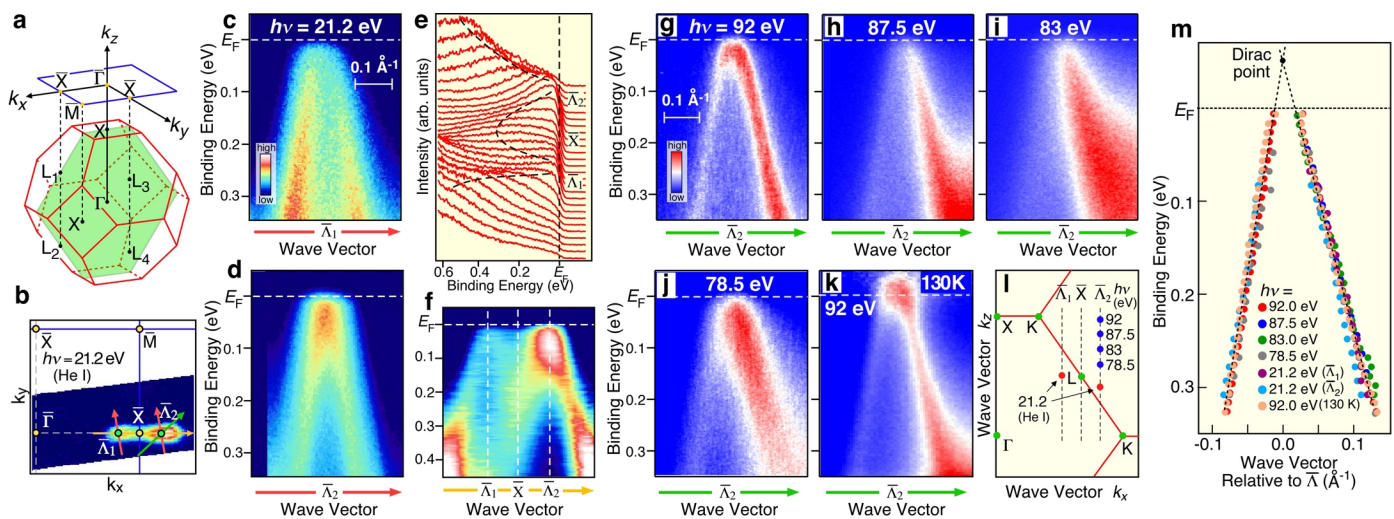

Fig. 1 Tanaka

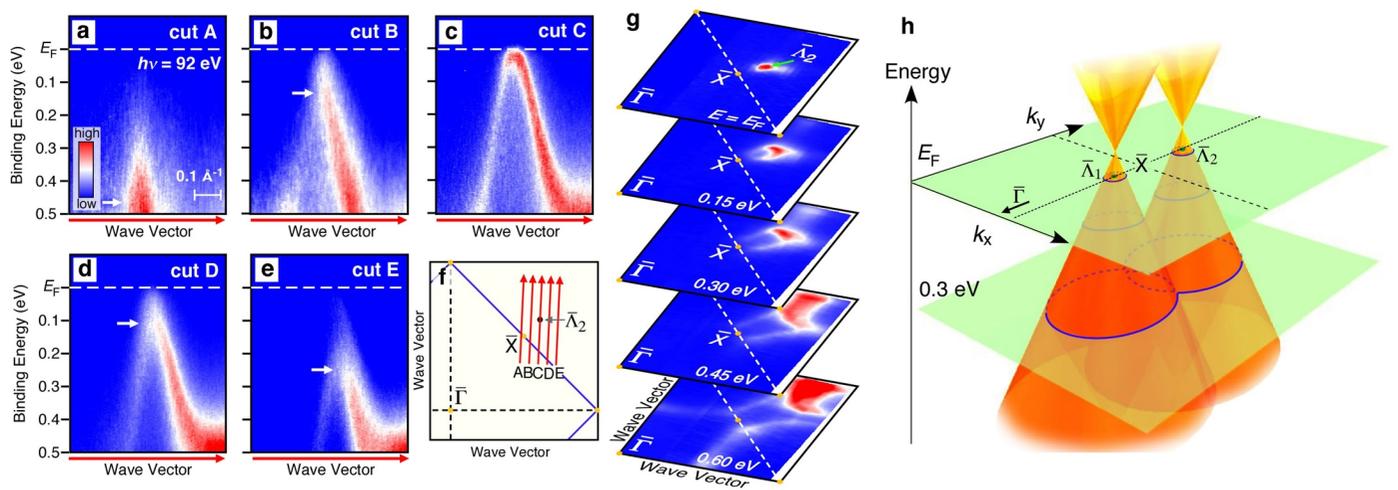

Fig. 2 Tanaka

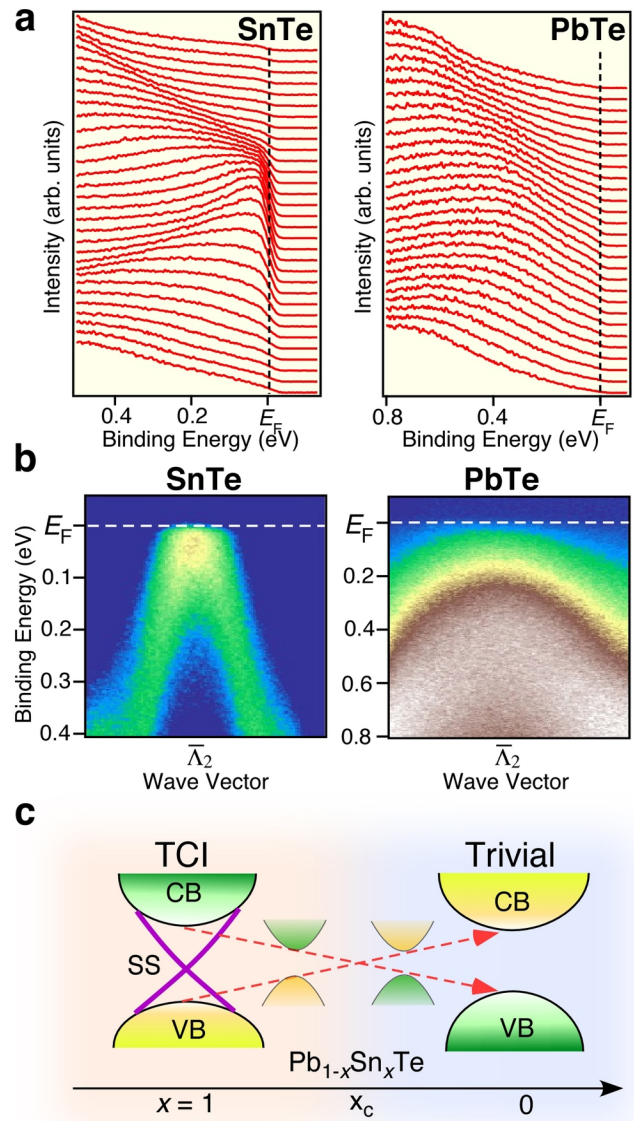

Fig. 3 Tanaka